\documentclass[aps,prb,reprint,superscriptaddress,amssymb,showpacs,showkeys,floatfix]{revtex4-1}

\usepackage{graphicx,amsmath,bm}

\newcommand{\uu}[1]{\ensuremath\,\mathrm{#1}}

\begin{document}

\date{\today}
\title{Trends in the elastic response of binary early transition metal nitrides}

\author{David Holec}
\email{david.holec@unileoben.ac.at}
\affiliation{Department of Physical Metallurgy and Materials Testing, Montanuniversit\"at Leoben, Franz-Josef-Strasse 18, A-8700 Leoben, Austria}
\author{Martin Fri\'ak}
\affiliation{Max-Planck-Institut f\"ur Eisenforschung GmbH, Max-Planck-Strasse 1, D-40237 D\"usseldorf, Germany}
\author{J\"org Neugebauer}
\affiliation{Max-Planck-Institut f\"ur Eisenforschung GmbH, Max-Planck-Strasse 1, D-40237 D\"usseldorf, Germany}
\author{Paul H.~Mayrhofer}
\affiliation{Department of Physical Metallurgy and Materials Testing, Montanuniversit\"at Leoben, Franz-Josef-Strasse 18, A-8700 Leoben, Austria}

\begin{abstract}
Motivated by an increasing demand for coherent data that can be used for selecting materials with properties tailored for specific application requirements, we studied elastic response of nine binary early transition metal nitrides (ScN, TiN, VN, YN, ZrN, NbN, LaN, HfN, and TaN) and AlN. In particular, single crystal elastic constants, Young's modulus in different crystallographic directions, polycrystalline values of shear and Young's moduli, and the elastic anisotropy factor were calculated. Additionally, we provide estimates of the third order elastic constants for the ten binary nitrides.
\end{abstract}



\maketitle

\section{Introduction}

Nitride compounds are a prominent class of materials with applications spanning from protective hard coatings (mostly transition metal nitrides, TMNs, of the IIIB-VIB group but also e.g. BN or SiN)\cite{PalDey2003,Mayrhofer2006}, to optoelectronic devices (mostly IIIA and VA groups, but also e.g. ScN or TiN)\cite{Jain2000}, to potential hydrogen storage materials such as Li$_3$N\cite{Chen2002}.

When it comes to their superior mechanical properties such as high hardness and Young's modulus, TMNs, and in particular TiN are often the industrial material of choice for surface coating of e.g. cutting tools. The properties of simple binary compounds can be successfully enhanced by forming metastable alloys, e.g. Ti$_{1-x}$Al$_x$N \cite{PalDey2003}, which at higher temperatures age-hardens before decomposing into its stable constituents\cite{Mayrhofer2003}. Recent studies have shown beneficial effects (such as increased oxidation resistance or retardation of the final decomposition step to higher temperatures) of additional alloying elements in Ti$_{1-x}$Al$_x$N\cite{Moser2007,Mayrhofer2010,Mayrhofer2003,Chen2011} and other systems\cite{Rovere2010a}. Another approach how to improve material properties is via multilayer design, where individual layers are typically simple binary or ternary systems\cite{Holleck1990,Friesen1991,Paulitsch2010}.

A modern way designing new and improve current materials is to combine experiment with modelling. For simple and/or small systems, quantum mechanical first principle approaches can be used. However, these are typically limited to several hundreds of atoms and, e.g. multilayers, crack propagation or nanoindentor tip--layer contact become difficult topics to handle. Here, the continuum mechanics employing finite element method (FEM) proves to be a successful tool\cite{Mackerle2005,Holmberg1998,Walter2009}. A key prerequisite to perform FEM calculations is the knowledge of the elastic properties of the studied materials, which are not always experimentally available (e.g., because some phases are stable only in the multilayer arrangement, but not as a bulk material). In such cases the elastic constants can be provided by the first principle calculations.

The literature on first principle calculations of early TMNs (group IIIB-VB) is vast. The main focus of those papers is the electronic structure and related material chemistry problems, while the calculation of the elastic properties is often only a minor part of the results. Additionally, a lot of those reports focus only on one (or a few) systems. There are some exhaustive reports on the chemical trends in the early TMNs\cite{Haglund1991,Haglund1993,Eck1999,Stampfl2001,Wu2005,Zhao2008}, but apart from Refs.~\onlinecite{Wu2005,Zhao2008} they do not discuss the elastic properties. In addition, there are some discrepancies between reported values (e.g., $C_{11}$ for ZrN between $304\uu{GPa}$\cite{Cheng2004} and $616\uu{GPa}$\cite{Wu2005}) which are worth cross-checking.

The aim of the present paper is to give a comprehensive overview on elastic properties of early transition metal nitrides and AlN, since these materials are or have the potential to be used as protective coatings\footnote{We deliberately leave our CrN for this report, since its magnetic configuration deserves a deeper analysis, see e.g. Refs.~\onlinecite{Alling2010,Alling2010a,Rivadulla2009}.}. In particular, we investigate ten binary systems, AlN, ScN, TiN, VN, YN, ZrN, NbN, LaN, HfN, and TaN. We focus on the cubic variant (B1, $Fm3m$, NaCl prototype), which is the stable configuration of all of them apart from AlN, NbN, and TaN being metastable in this configuration. The single crystal elastic constants are validated by several independent approaches as well as by a comparison with available theoretical and experimental data. Subsequently, we calculate directionally resolved Young's modulus, anisotropy factors and polycrystalline elastic properties of these compounds, and rationalise the trends in terms of their electronic structure and bonding.

\section{Methodology}
 
\subsection{Deformation modes}\label{sec:def_modes}

The linear elastic response of cubic materials is fully described by three independent components $c_{xxxx}$, $c_{xxyy}$, and $c_{xyxy}$ of the fourth rank tensor of the second order elastic constants (SOECs). It is convenient to represent this tensor with a $6\times6$ matrix 
\begin{equation}
C_{ij}=\begin{pmatrix} C_{11} & C_{12} & C_{12} & 0 & 0 & 0 \\ C_{12} & C_{11} & C_{12} & 0 & 0 & 0 \\ C_{12} & C_{12} & C_{11} & 0 & 0 & 0 \\ 0 & 0 & 0 & C_{44} & 0 & 0 \\ 0 & 0 & 0 & 0 & C_{44} & 0 \\ 0 & 0 & 0 & 0 & 0 & C_{44} \end{pmatrix}
\end{equation}
where $C_{11}=c_{xxxx}$, $C_{12}=c_{xxyy}$, and $C_{44}=C_{66}=c_{xyxy}$. Here we make use of the Voigt notation $xx\sim1$, $yy\sim2$, $zz\sim3$, $yz\sim4$, $xz\sim5$, and $xy\sim6$. An additional relationship links SOECs with the bulk modulus, $B$, \begin{equation}
  B=\frac13(C_{11}+2C_{12})\ .\label{bulkModulus}
\end{equation}
$B$ describes the elastic response of materials to volume changes, and it is obtained as a fitting parameter from the Birch-Murnaghan equation of state\cite{Birch1947}. Consequently, two other deformation modes are needed to obtain all independent components of the cubic elastic tensor.

The first pair consists of orthorhombic and monoclinic deformations. The orthorhombic mode results in a strain tensor
\begin{equation}
  \varepsilon_{\mathrm{orth}}(\delta)=\begin{pmatrix}
    \delta & 0 & 0 \\
    0 & -\delta & 0 \\
    0 & 0 & \frac{\delta^2}{1-\delta^2}
  \end{pmatrix}
\end{equation}
and the corresponding strain energy density, $U(\delta)$, is
\begin{equation}
  U_{\mathrm{orth}}(\delta)={\mathcal E}_{\mathrm{tot}}(\delta)-{\mathcal E}_{\mathrm{eq}}=(C_{11}-C_{12})\delta^2+{\cal O}(\delta^3)\ .
\end{equation}
Here, ${\mathcal E}_{\mathrm{tot}}(\delta)$ and ${\mathcal E}_{\mathrm{eq}}$ are the total energies per unit volume, corresponding to $\varepsilon_{\mathrm{orth}}(\delta)$ and $\varepsilon_{\mathrm{orth}}(0)$, respectively. A monoclinic deformation yielding a strain tensor
\begin{equation}
  \varepsilon_{\mathrm{mon}}(\delta)=\begin{pmatrix}
    0 & \frac12\delta & 0 \\
    \frac12\delta & 0 & 0\\
    0 & 0 & \frac{4}{4-\delta^2}
  \end{pmatrix}
\end{equation}
is used to evaluate the $C_{44}$ elastic constant from the corresponding strain energy density
\begin{equation}
  U_{\mathrm{mon}}(\delta)={\mathcal E}_{\mathrm{tot}}(\delta)-{\mathcal E}_{\mathrm{eq}}=\frac12C_{44}\delta^2+{\cal O}(\delta^3)\ .
\end{equation}
One should note that these two modes keep the unit cell volume constant.

The second pair of deformations, which is also often used, is a pair of tetragonal and triclinic distortion. The tetragonal deformation corresponds to a strain matrix
\begin{equation}
  \varepsilon_{\mathrm{tet}}(\delta)=\begin{pmatrix}
    -\frac12\delta & 0 & 0 \\
    0 & -\frac12\delta & 0 \\
    0 & 0 & \delta
  \end{pmatrix}
\end{equation}
producing a strain energy density
\begin{equation}
  U_{\mathrm{tet}}(\delta)={\mathcal E}_{\mathrm{tot}}(\delta)-{\mathcal E}_{\mathrm{eq}}=\frac34(C_{11}-C_{12})\delta^2+{\cal O}(\delta^3)\ .
\end{equation}
The $C_{44}$ elastic constants is obtained from a trigonal distortion with a strain tensor
\begin{equation}
  \varepsilon_{\mathrm{tri}}(\delta)=\begin{pmatrix}
    0 & \delta & 0 \\
    \delta & 0 & 0\\
    0 & 0 & 0
  \end{pmatrix}
\end{equation}
and a strain energy density
\begin{equation}
  U_{\mathrm{tri}}(\delta)={\mathcal E}_{\mathrm{tot}}(\delta)-{\mathcal E}_{\mathrm{eq}}=2C_{44}\delta^2+{\cal O}(\delta^3)\ .
\end{equation}
These two deformations are volume non-conserving.

Recently, Zhao \textit{et al.}\cite{Zhao2007} proposed a set of six deformation matrices which allow for estimation of second and third order\footnote{There are in total six independent third order elastic constants for materials with the cubic symmetry.} elastic constants of cubic materials at the same time. These are:
\begin{gather}
  A1 = \begin{pmatrix} \delta & 0 & 0 \\ 0 & 0 & 0\\ 0 & 0 & 0 \end{pmatrix}\ ,\quad A2 = \begin{pmatrix} \delta & 0 & 0 \\ 0 & \delta & 0\\ 0 & 0 & 0 \end{pmatrix}\ ,\notag\\ 
  A3 = \begin{pmatrix} \delta & 0 & 0 \\ 0 & \delta & 0\\ 0 & 0 & \delta \end{pmatrix}\ ,\quad A4 = \begin{pmatrix} \delta & 0 & 0 \\ 0 & 0 & \delta\\ 0 & \delta & 0 \end{pmatrix}\ , \\
  A5 = \begin{pmatrix} \delta & \delta & 0 \\ \delta & 0 & 0\\ 0 & 0 & 0 \end{pmatrix}\ ,\quad A6 = \begin{pmatrix} 0 & \delta & \delta \\ \delta & 0 & \delta\\ \delta & \delta & 0 \end{pmatrix}\ .\notag
\end{gather}
The strain energy density in these cases is
\begin{equation}
  U_{A}(\delta)={\mathcal E}_{\mathrm{tot}}(\delta)-{\mathcal E}_{\mathrm{eq}}={\cal A}\delta^2+{\cal B}\delta^3+{\cal O}(\delta^4) \label{eq:TEOC}
\end{equation}
where the coefficients ${\cal A}$ and ${\cal B}$ are specific combinations of $C_{ij}$ and $C_{ijk}$ as given in Table~\ref{tab:combinations}.

\begin{table}
  \begin{ruledtabular}
  \begin{tabular}{lcc}
	 & $\cal A$ & $\cal B$ \\\hline
    $A1$ & $\frac12 C_{11}$ & $\frac16 C_{111}$ \\
    $A2$ & $C_{11}+C_{12}$ & $\frac13 C_{111}+C+{112}$ \\
    $A3$ & $\frac32 C_{11}+3 C_{12}$ & $\frac12 C_{111}+3C_{112}+C_{123}$ \\
    $A4$ & $\frac12 C_{11}+2 C_{44}$ & $\frac16 C_{111}+2C_{144}$ \\
    $A5$ & $\frac12 C_{11}+2 C_{44}$ & $\frac16 C_{111}+2C_{166}$ \\
    $A6$ & $6 C_{44}$ & $6 C_{456}$ \\
  \end{tabular}
  \end{ruledtabular}

  \caption{Coefficients from Eq.~\ref{eq:TEOC} for various deformation matrices $A1$--$A6$.}\label{tab:combinations}
\end{table}

\subsection{Calculation details}

Quantum mechanical calculations employing density functional theory (DFT)\cite{Hohenberg1964,Kohn1965} were carried out using Vienna Ab initio Simulation Package\cite{Kresse1993,Kresse1996}. Projector augmented-wave pseudopotentials\cite{Kresse1999} together with the generalised gradient approximation (GGA), as parametrised by Wang and Perdew\cite{Wang1991}, for the exchange and correlation potential are used. The plane-wave cut-off energies and the $\bm{k}$-vector samplings of the Brillouin zone were carefully checked to provide a total energy accuracy in the order of $1\uu{meV/at}$ or better. They are listed in Table~\ref{tab:calc_details} together with the used pseudopotentials; the suffices \_sv and \_pv refer to the exact valence configuration taking into account explicitely also the $s$ and $p$ closed-shell electrons, respectivelly.

\begin{table}
  \begin{ruledtabular}
  \begin{tabular}{lcc}
    pseudopotential & $E_{\mathrm{cut}}\uu[eV]$ & $\bm{k}$-point sampling\\\hline\hline
    Al     & 800 & 5$\times$5$\times$5\\\hline
    Sc\_sv & 800 & 7$\times$7$\times$7\\
    Ti\_pv & 400 & 17$\times$17$\times$17\\
    V\_pv  & 800 & 11$\times$11$\times$11\\\hline
    Y\_sv  & 800 & 7$\times$7$\times$7\\
    Zr\_sv & 700 & 13$\times$13$\times$13\\
    Nb\_sv & 600 & 15$\times$15$\times$15\\\hline
    La     & 800 & 7$\times$7$\times$7\\
    Hf\_pv & 700 & 11$\times$11$\times$11\\
    Ta\_pv & 600 & 15$\times$15$\times$15
  \end{tabular}
  \end{ruledtabular}
  \caption{An overview PAW-GGA pseudopotentials, plane wave cut-off energies, and the Monkhorst-Pack sampling of the Brillouin zone used in this study.}\label{tab:calc_details}
\end{table}

\section{Results and Discussion}

\subsection{Equilibrium properties}

The optimised lattice constants, $a$, formation energies, $E_f$, and mass densities, $\rho$, are summarised in Table~\ref{tab:equilibrium_properties}. Since there have been a vast number of publications on experimental as well as calculated equilibrium structure parameters of these early TMN compounds (see e.g., Refs.~\onlinecite{Eck1999, Haglund1991, FernandezGuillermet1992, Stampfl2001, Zhao2008, Rovere2010}, and references therein), we limit the comparison of the here calculated lattice parameters to the experimental data from Ref.~\onlinecite{JCPDF}. The calculated lattice constants are, as expected for GGA, slightly larger than the experimental values. The error is smaller then 1\% except for the case of TaN, where a deviation of about 1.5\% is obtained. This is likely to be related to the fact that cubic TaN is metastable and prefers N deficient configurations resulting in a significant decrease of the lattice parameter with respect to a stoichiometric configuration\cite{Rachbauer2010}. 

\begin{table}
  \begin{ruledtabular}
  \begin{tabular}{l|cc|c|cc}
        & $a\uu{[\AA]}$ & $a_{\mathrm{exp}}\uu{[\AA]}$ & $E_f\uu{[eV/at]}$ & $\rho\uu{[g/cm^3]}$ & $\rho_{\mathrm{exp}}\uu{[g/cm^3]}$ \\\hline\hline
    AlN & 4.069 & 4.045 & $-2.285$ & 4.04 &              \\\hline
    ScN & 4.516 & 4.440 & $-2.958$ & 4.25 &              \\
    TiN & 4.253 & 4.241 & $-2.752$ & 5.34 & 5.40$^\star$ \\
    VN  & 4.127 & 4.139 & $-1.998$ & 6.14 & 6.13, 6.0$^\star$         \\\hline
    YN  & 4.917 & 4.894 & $-2.737$ & 5.75 &              \\
    ZrN & 4.618 & 4.578 & $-2.716$ & 7.10 & 7.32$^\star$ \\
    NbN & 4.427 & 4.389 & $-2.001$ & 8.19 & 8.47, 7.3$^\star$         \\\hline
    LaN & 5.306 & 5.293 & $-2.350$ & 6.80 & 6.73         \\
    HfN & 4.538 & 4.525 & $-2.783$ & 13.68 & 13.80, 13.8$^\star$       \\
    TaN & 4.426 & 4.358 & $-1.869$ & 14.94 & 13.70, 14.3$^\star$     
  \end{tabular}
  \end{ruledtabular}

  \caption{Calculated lattice constants, $a$, formation energy, $E_f$, and mass density, $\rho$. The experimental lattice constants $a_{\mathrm{exp}}$ are taken from Ref.~\onlinecite{JCPDF}, the experimental values of density $\rho_{\mathrm{exp}}$ are from Ref.~\onlinecite{Chem} and those marked with an asterisks are from Ref.~\onlinecite{Pierson1996}.}\label{tab:equilibrium_properties}.
\end{table}

The trends in the energy of formation, i.e. less negative values as one moves from the IIIB to the VB group, as well as the absolute numbers agree well with those presented by Rovere \textit{et al.}\cite{Rovere2010}.

Lastly, from the calculated lattice parameters (equilibrium volume) and the atomic weights we computed mass densities. Again, apart from the TaN case, where the under-stoichiometry of the experimental compound is likely to play a role, the agreement with experimental values is satisfactory.

\subsection{Single crystal elastic constants}

\begin{figure*}
  \includegraphics{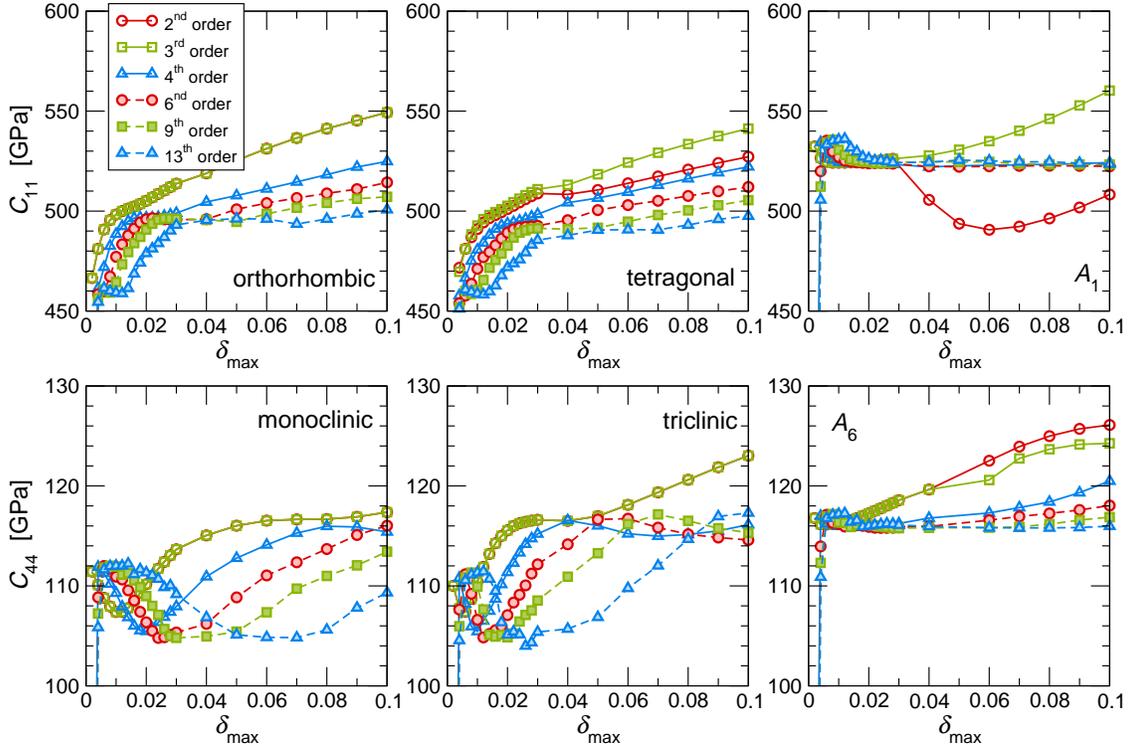}

  \caption{(Colour on-line) Dependence of the ZrN elastic constants on the range, $-\delta_{\max}\leq\delta\leq\delta_{\max}$, of deformation taken into account, and the order of the fitting polynomial. The upper and lower row correspond to $C_{11}$ and $C_{44}$, respectively, as obtained from various approaches described in Sec.~\ref{sec:def_modes}.}\label{fig:plateau}
\end{figure*}

When calculating the single crystal elastic constants as described in Sec.~\ref{sec:def_modes}, one should check how the fitted $C_{ij}$s depend on the maximum deformation, $\delta_{\max}$, i.e., on the range of deformations applied to the unit cell. When $\delta_{\max}$ is too small, the accuracy of $C_{ij}$s is likely to be influenced by the numerical inaccuracies of the DFT calculations, while non-linear elastic (and perhaps also plastic) effects are no longer negligible for too large $\delta_{\max}$\cite{Zhao2007}. To illustrate this behaviour, we plot in Fig.~\ref{fig:plateau} the $C_{11}$ and $C_{44}$ elastic constants of ZrN  as a function of $\delta_{\max}$ and the order of the fitting polynomial. It follows, that with increasing order of the fitting polynomial, the plateau region where the specific elastic constant is independent of $\delta_{\max}$, enlarges. At the same time, the onset of the plateau shifts to higher values $\delta_{\max}$. The reason is that a high fitting order leads to an over-fitting of the too few data points for a small $\delta_{\max}$. In extreme cases, such over-fitting may lead to incorrect plateaus, as shown e.g. for $C_{44}$ (monoclinic deformation) using a $13^{\mathrm{th}}$ order fitting polynomial.

In general, the combination of tetragonal and trigonal deformations gives more robust results for the early TMN in the cubic structure then the combination of orthorhombic and monoclinic deformations. Nevertheless, the most robust results in terms of plateau values scatter and the dependence on the order of the fitting polynomial, were obtained when employing the deformation matrices $A1$ and $A6$ (see Fig.~\ref{fig:plateau}).

Recently, Udyansky \textit{et al.}\cite{Udyansky2011} showed that the elastic constants of $\alpha$-Fe are also highly sensitive to the value of the smearing parameter, $\sigma$, in the Methfessel-Paxton scheme. We have therefore checked the convergence of $C_{ij}$s also with respect to $\sigma$. It turns out that the cubic early TMNs are not hugely sensitive to $\sigma$, but in some cases, e.g., ZrN or NbN, the elastic constant values change by up to 5\% when sigma is increased from $0.02\uu{eV}$ to $0.8\uu{eV}$. Nevertheless, these variations typically take place only for small values of $\sigma$, and a converged behaviour is obtained for $\sigma\approx0.6\uu{eV}$.

\begin{table}
  \begin{ruledtabular}
  \begin{tabular}{lllllll}
	&  \multicolumn{2}{c}{$C_{11}\uu{[GPa]}$} & \multicolumn{2}{c}{$C_{12}\uu{[GPa]}$} & \multicolumn{2}{c}{$C_{44}\uu{[GPa]}$} \\\hline
	& o:   420 & c: 423\footnotemark[1] & o:   166 & c: 167\footnotemark[1] & o:   309 & c: 306\footnotemark[1] \\
    AlN & t:   421 & c: 379\footnotemark[2] & t:   168 & c: 201\footnotemark[2] & t:   308 & c: 196\footnotemark[2] \\
	& $A$: \textbf{418} &     & $A$: \textbf{169} &     & $A$: \textbf{308} &     \\\hline
	& o:   390 & c: 498\footnotemark[3] & o:   105 & c: 52\footnotemark[3] & o:   166 & c: 169\footnotemark[3] \\
    ScN & t:   388 & c: 299\footnotemark[4] & t:   105 & c: 128\footnotemark[4] & t:   166 & c: 120\footnotemark[4]\\
	& $A$: \textbf{388} & c: 381\footnotemark[5] & $A$: \textbf{106} & c: 105\footnotemark[5] & $A$: \textbf{166} & c: 164\footnotemark[5]\\\hline
	& o:   560 & c: 516\footnotemark[6] & o:   135 & c: 129\footnotemark[6] & o:   163 & c: 132\footnotemark[6] \\
    TiN & t:   577 & c: 610\footnotemark[1] & t:   129 & c: 137\footnotemark[1] & t:   161 & c: 158\footnotemark[1] \\
	& $A$: \textbf{575} & e: \textit{625}\footnotemark[7] & $A$: \textbf{130} & e: \textit{165}\footnotemark[7] & $A$: \textbf{163} & e: \textit{163}\footnotemark[7] \\\hline
	& o:   660  & c: 738\footnotemark[8] & o:   174  & c: 186\footnotemark[8] & o:   118 & c: 119\footnotemark[8] \\
    VN  & t:   658 &     & t:   172 &     & t:   118 &     \\
	& $A$: \textbf{660} & e: \textit{533}\footnotemark[7] & $A$: \textbf{144} & e: \textit{135}\footnotemark[7] & $A$: \textbf{120} & e: \textit{133}\footnotemark[7] \\\hline
	& o:   318 &     & o:    81 &     & o:   124 &     \\
    YN  & t:   318 &     &  t:   81 &     & t:   124 &     \\
	& $A$: \textbf{318} &     &  $A$: \textbf{81} &     & $A$: \textbf{124} &     \\\hline
	& o:   495 & c: 616\footnotemark[9] &  o:  128 & c: 117\footnotemark[9] & o:   116 & c: 130\footnotemark[9] \\
    ZrN & t:   492 & c: 304\footnotemark[12] & t:   126 & c: 114\footnotemark[12] & t:   116 & c: 511\footnotemark[12] \\
	& $A$: \textbf{523} & e: \textit{471}\footnotemark[10] & $A$: \textbf{111} & e: \textit{88}\footnotemark[10] & $A$: \textbf{116} & e: \textit{138}\footnotemark[10] \\\hline
	& o:   --  & c: 739\footnotemark[9] & o:   --  & c: 161\footnotemark[9] & o:   --  & c: 75\footnotemark[9] \\
    NbN & t:   --  & e: \textit{556}\footnotemark[7] & t:   --  & e: \textit{152}\footnotemark[7] & t:   --  & e: \textit{125}\footnotemark[7] \\
	& $A$: \textbf{649} & e: \textit{608}\footnotemark[10] & $A$: \textbf{136} & e: \textit{134}\footnotemark[10] & $A$: \textbf{80}  & e: \textit{117}\footnotemark[10] \\\hline
	& o:   201 & c: 221\footnotemark[11] & o:   84  & c: 62\footnotemark[11] & o:   71  & c: 75\footnotemark[11] \\
    LaN & t:   201 & c: 213\footnotemark[13] & t:   83  & c: 84\footnotemark[13] & t:   71  & c: 71\footnotemark[13] \\
	& $A$: \textbf{198} &     & $A$: \textbf{86}  &     & $A$: \textbf{71}  &     \\\hline
	& o:   575 & c: 694\footnotemark[9] & o:   120 & c: 112\footnotemark[9] & o:   117 & c: 135\footnotemark[9] \\
    HfN & t:   --  & c: 628\footnotemark[11] & t:   --  & c: 95\footnotemark[11] & t:   --  & c: 105\footnotemark[11] \\
	& $A$: \textbf{588} & e: \textit{679}\footnotemark[10] & $A$: \textbf{113} & e: \textit{119}\footnotemark[10] & $A$: \textbf{120} & e: \textit{150}\footnotemark[10] \\\hline
	& o:   --  & c: 783\footnotemark[9] & o:   --  & c: 167\footnotemark[9] & o:   --  & c: 20\footnotemark[9] \\
    TaN & t:   --  & c: 881\footnotemark[11] & t:   --  & c: 122\footnotemark[11] & t:   57  & c: 74\footnotemark[11] \\
	& $A$: \textbf{715} &     & $A$: \textbf{138} &     & $A$: \textbf{60}  &     \\
  \end{tabular}
  \end{ruledtabular}
  \footnotetext[1]{Ref.~\onlinecite{Wang2010}, GGA}
  \footnotetext[2]{Ref.~\onlinecite{Saib2006}, GGA}
  \footnotetext[3]{Ref.~\onlinecite{Pandit2011}, LDA}
  \footnotetext[4]{Ref.~\onlinecite{Oussaifi2007}, GGA}
  \footnotetext[5]{Ref.~\onlinecite{Feng2010}, GGA}
  \footnotetext[6]{Ref.~\onlinecite{Zhang2001}, LDA}
  \footnotetext[7]{Ref.~\onlinecite{Kim1992}, exp.}
  \footnotetext[8]{Ref.~\onlinecite{Wolf1999}, LDA}
  \footnotetext[9]{Ref.~\onlinecite{Wu2005}, LDA}
  \footnotetext[10]{Ref.~\onlinecite{Chen2005}, exp.}
  \footnotetext[11]{Ref.~\onlinecite{Zhao2008}, GGA}
  \footnotetext[12]{Ref.~\onlinecite{Cheng2004}, GGA}
  \footnotetext[13]{Ref.~\onlinecite{Ciftci2008}, GGA}
  \caption{Single crystal elastic constants. ``o'', ``t'', and ``A'' stand from values calculated by orthorhombic and monoclinic, tetragonal and trigonal, and using the $A1$ and $A6$ deformation modes, respectively. ``c'' and ``e'' stand for calculated and experimental data from the literature, respectively. The results in bold are used for the further analysis of polycrystalline elastic properties.}\label{tab:elastic_constants}
\end{table}

\begin{figure}
  \includegraphics{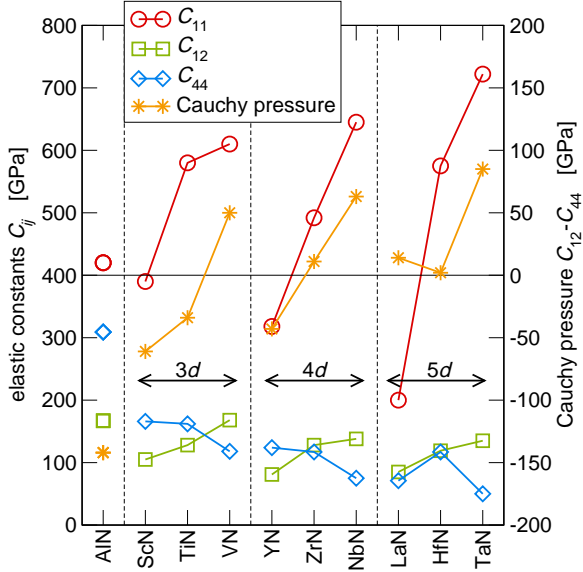}
  \caption{(Colour on-line) Single crystal elastic constants, $C_{11}$, $C_{12}$, and $C_{44}$, and resulting Cauchy pressure, $C_{12}-C_{44}$, of the compounds investigated here.}\label{fig:elastic_constants}
\end{figure}

The single crystal elastic constants are summarised in Table~\ref{tab:elastic_constants}. When possible to evaluate, we give the $C_{ij}$s based on all three methods described here (i.e., orthorhombic+monoclinic, tetragonal+trigonal, and $A1$+$A6$ deformation modes). Since the $A1$+$A6$ deformation modes were the only ones to provide well converged results for all ten binary nitrides, we show them in Fig.~\ref{fig:elastic_constants} and we will use them in the following analysis for consistency.

A comparison with previous DFT-GGA literature data\cite{Wang2010, Saib2006, Oussaifi2007, Feng2010, Zhao2008, Cheng2004, Ciftci2008} yields, apart from a few exceptions, a good agreement with our results. The local density approximation (LDA) based $C_{11}$ elastic constants from the literature\cite{Pandit2011, Wolf1999, Wu2005} are higher than our GGA-based data. This is a consequence of over- and under-binding of LDA and GGA, respectively, resulting in too small lattice constants and consequently too hard elastic constants in LDA. Finally, although many of the here calculated elastic constants agree well with the available experimental data, in a few cases the discrepancy is as large as 20\% (e.g. $C_{11}$ of VN).

In all cases, the obtained elastic constants fulfil the stability criteria for cubic crystals
\begin{equation}
  C_{44}>0\ ,\quad C_{11}>|C_{12}|\ ,\quad C_{11}+2C_{12}>0\ .
\end{equation}
The elastic constant $C_{11}$ is significantly stiffer than the other two elastic constant. Within each row the $C_{11}$ and $C_{12}$ elastic constants monotonically increase with increasing atomic number at the same time, $C_{11}$ decreases from Sc to Y to La (isovalent IIIB group) while it increases from V to Nb to Ta (isovalent VB group). It has been suggested in the literature\cite{Chen2003,Pettifor1992}, that a negative Cauchy pressure $C_{12}-C_{44}<0$ corresponds to more directional bonding while positive values indicate predominant metallic bonding. Indeed, the calculated Cauchy pressure is most negative for AlN in which a significantly larger charge transfer from cation to anion takes place as e.g., in TiN\cite{Holec2011}. The Cauchy pressure increases to positive values with increasing number of valence electrons within each periodic table row, as those contribute mainly to the metal--metal $d$--$d$ interactions\cite{Stampfl2001,Rachbauer2011}. These trends may be used in the materials selection process to realize specific requirements.

\subsection{Directional Young's modulus}

The Young's modulus, $E$, in a certain direction, $\bm{\xi}$, is defined as the ratio of longitudinal stress to longitudinal strain in this direction. The elastic compliances, $S_{ij}$, are in the case of cubic crystals the solution of the following set of equations\cite{Nye1957}:
\begin{subequations}
\begin{gather}
  C_{11}=\frac{S_{11}+S_{12}}{(S_{11}-S_{12})(S_{11}+2S_{12})}\ ,\\
  C_{12}=\frac{-S_{12}}{(S_{11}-S_{12})(S_{11}+2S_{12})}\ ,\\
  C_{44}=\frac{1}{S_{44}}\ . 
\end{gather}
\end{subequations}
 For a cubic crystals $E_{\bm{\xi}}$ then reads\cite{Nye1957}
\begin{equation}
  \frac1{E_{\bm{\xi}}}=S_{11}-2\left(S_{11}-S_{12}-\frac12S_{44}\right)(l_1^2l_2^2+l_2^2l_3^2+l_1^2l_3^2)
\end{equation}
where $l_1$, $l_2$, and $l_3$ are the directional cosines of $\bm{\xi}$. For $\langle100\rangle$, $\langle110\rangle$, and $\langle111\rangle$ directions this becomes
\begin{subequations}
\begin{gather}
  E_{\langle100\rangle}=1/S_{11}\ ,\\
  E_{\langle110\rangle}=1\left/\left(S_{11}-\frac12\left(S_{11}-S_{12}-\frac12S_{44}\right)\right)\right.\ ,\\
  E_{\langle111\rangle}=1\left/\left(S_{11}-\frac23\left(S_{11}-S_{12}-\frac12S_{44}\right)\right)\right.\ .
\end{gather}
\end{subequations}

\begin{figure}
  \includegraphics{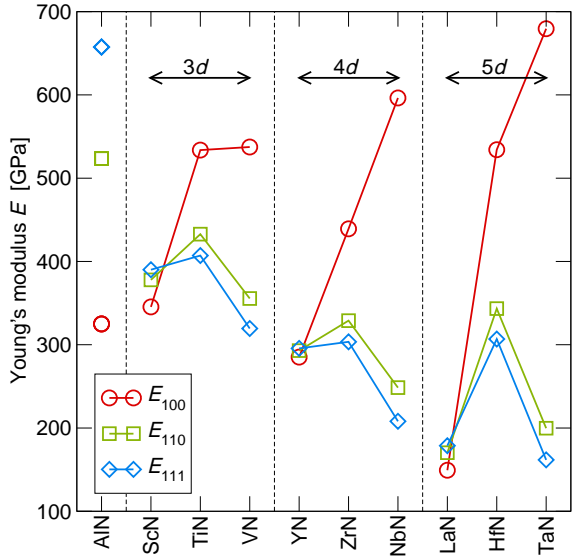}
  \caption{(Colour on-line) Young's modulus in the $\langle100\rangle$, $\langle110\rangle$, and $\langle111\rangle$ directions.}\label{fig:directional_E}
\end{figure}

The results are plotted in Fig.~\ref{fig:directional_E}. The Young's modulus in $\langle100\rangle$ follows mostly the same trend as $C_{11}$, since the $C_{11}$ elastic constant has the strongest contribution to $E_{\langle100\rangle}$. There is a considerable difference between the semiconducting compounds AlN, ScN, YN, and LaN, in which the $\langle100\rangle$ direction becomes the softest, and the metallic TiN, VN, ZrN, NbN, HfN, and TaN, where the $\langle100\rangle$ direction is clearly the strongest. In addition, the Young's modulus of AlN in the $\langle111\rangle$ direction is more than 1.5-times larger than in any other of the here investigated TMNs. This is mainly caused  by the high value of $C_{44}$ of AlN, suggesting that AlN is much stronger in shear deformation than the other TMN.

\begin{figure}
  \includegraphics{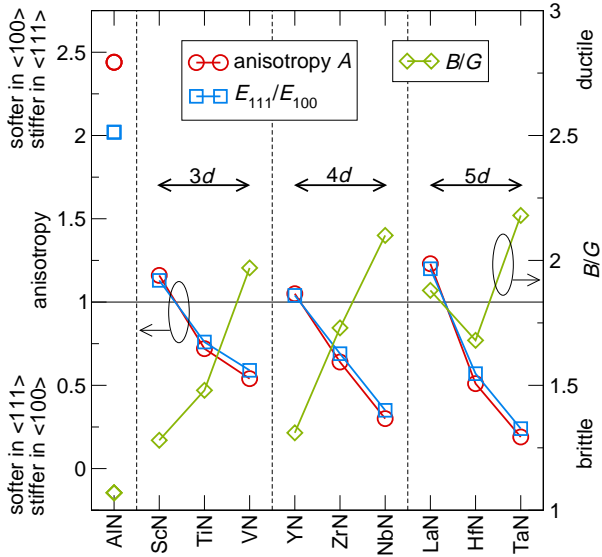}
  \caption{(Colour online) Zener's anisotropy ratio, $E_{\langle111\rangle}/E_{\langle100\rangle}$, and $B/G$ values for the ten nitrides studied here.}\label{fig:anisotropy}
\end{figure}

\begin{figure}
  \includegraphics[width=8cm]{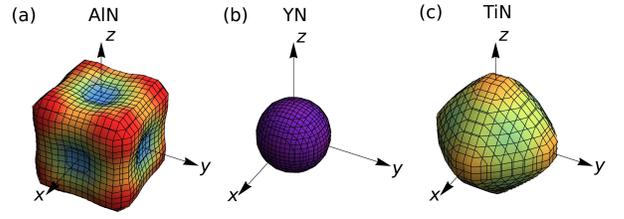}
  \caption{(Colour on-line) 3D representation of the directional dependence of Young's modulus for (a)~AlN, (b)~YN, and (c)~TiN.}\label{fig:directional_E_3D}
\end{figure}

To quantify the anisotropy, we employ the Zener's anisotropy ratio\cite{Ranganathan2008}, $A$, defined as
\begin{equation}
  A=\frac{2C_{44}}{C_{11}-C_{12}}\ .\label{eq:anisotropy}
\end{equation}
The results, together with the ratio $E_{\langle111\rangle}/E_{\langle100\rangle}$ which provides similar information, are shown in Fig.~\ref{fig:anisotropy}. The results suggest that AlN is clearly stiffer in the $\langle111\rangle$ than in the $\langle100\rangle$ direction. The opposite result is obtained for the group IVB and VB TMN where the $\langle100\rangle$ direction is the stiffest. The group IIIB semiconducting TMN exhibit values of both, $A$ and $E_{\langle111\rangle}/E_{\langle100\rangle}$, very close to 1. This implies that their elastic behaviour is almost isotropic. The most isotropic response is predicted for YN with $A=1.05$. The (an)isotropy of the Young's modulus is visualised in Fig.~\ref{fig:directional_E_3D}. Fig.~\ref{fig:directional_E_3D}b demonstrates the isotropic elastic response of YN, while the comparison of Figs.~\ref{fig:directional_E_3D}a and c shows the qualitatively different elastic response of AlN and TiN.

Some insight into these trends can be gained from considering the differences in bonding. The bonds in AlN are strongly ionic\cite{Holec2011} while the TMNs contain a significant part of the covalent bonding\cite{Stampfl2001}. Since the covalent bond is stronger than the ionic, this can rationalise why AlN has the smallest value of $E_{\langle100\rangle}$. When going from group IIIB to VB elements within each row, the extra electrons fill the bonding metal--metal orbitals, while the hybridised $sp^3d^2$ states move to lower energies\cite{Stampfl2001,Rachbauer2011}. This can be interpreted as strengthening of the hybridised $sp^3d^2$ bonds which are oriented along the $\langle100\rangle$ directions. As for the high $C_{44}$ elastic constant of AlN, one may argue that since there are no $d$ electrons available to form the metal-metal bonds in the $\langle110\rangle$ directions (as it is the case for the IVB and VB group elements), upon a shear deformation a significantly increased repulsion between Al-Al and N-N ions as they get closer occurs, which causes the high value of $E_{\langle111\rangle}$.

\subsection{Polycrystalline properties}

Several models exist which assess the isotropic polycrystalline elastic properties using the anisotropic single crystal elastic constants of a given material. Voigt's approach\cite{Voigt1928} of constant strains in all grains yields the upper limit, $G_V$ and $E_V$, to the polycrystalline shear and Young's moduli, respectively. On the other hand, Reuss\cite{Reuss1929} proposed to apply constant stresses in all grains, which yields  lower limits $G_R$ and $E_R$. Taking $B_V=B_R=B$, where the bulk modulus, $B$, is obtained from the Birch-Murnaghan equation of state\cite{Birch1947}, one gets
\begin{gather}
  G_V=\frac{C_{11}-C_{12}+3C_{44}}5\ ,\label{eq:ShearModulusVoigt}\\
  G_R=\frac5{4(S_{11}-S_{12})+3S_{44}}\ ,\\
  E_{\alpha}=\frac{9BG_{\alpha}}{3B+G_{\alpha}}\ ,\ \alpha=V,\,R\ .\label{eq:YoungsModulus}
\end{gather}
Finally, Hershey\cite{Hershey1954} derived an equation for self-consistently calculating the shear modulus, $G_H$. In this approach, $G_H$ is the real positive root of the following fourth order polynomial
\begin{widetext}
\begin{multline}
  64G_H^4+
  16(4C_{11}+5C_{12})G_H^3+
  \left(3(C_{11}+2C_{12})(5C_{11}+4C_{12})-8(7C_{11}-4C_{12})C_{44}\right)G_H^2-\\-
  (29C_{11}-20C_{12})(C_{11}+2C_{12})C_{44}G_H-
  3(C_{11}+2C_{12})^2(C_{11}-C_{12})C_{44}=0\ .
\end{multline}
This equation can be simplified by dividing it with $(3 C_{11} + 6 C_{12} + 8 G)$ to a third order polynomial \cite{Lubarda1997} with the same positive real root
\begin{equation}
  8G^3+(5C_{11}+4C_{12})G^2-C_{44}(7C_{11}-4C_{12})G-C_{44}(C_{11}-C_{12})(C_{11}+2C_{12})=0\ .
\end{equation}
\end{widetext}
Subsequently, Eq.~\ref{eq:YoungsModulus} is used to estimate the Young's modulus within Hershey's approach.

\begin{figure}
  \includegraphics{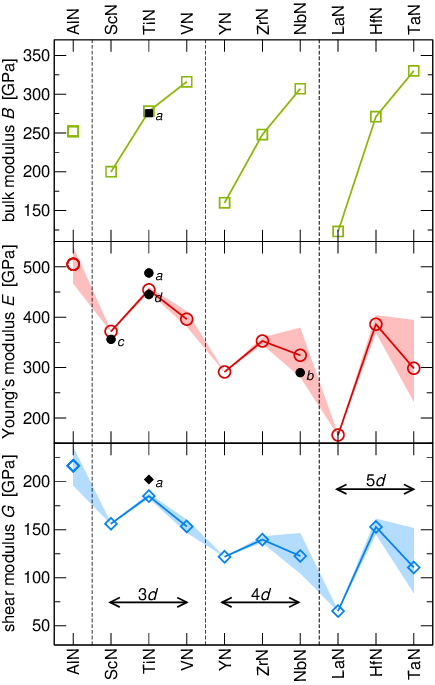}
  \footnotetext[1]{Ref.~\onlinecite{Chen2010}}
  \footnotetext[2]{Ref.~\onlinecite{Benkahoul2004}}
  \footnotetext[3]{Ref.~\onlinecite{Gall1998}}
  \footnotetext[4]{Ref.~\onlinecite{Ljungcrantz1996}}
  \caption{(Colour on-line) Calculated bulk modulus, $B$, and polycrystalline Young's, $E$, and shear, $G$, moduli. The shaded area correspond to the Reuss's and Voigt's limit cases. The full symbols denote experimental values from literature.}\label{fig:polycrystalline}
\end{figure}

The thus calculated polycrystalline elastic constants are shown in Fig.~\ref{fig:polycrystalline} together with the bulk modulus. They fit well with the few accessible experimental data-points included (black full symbols). The trends in  $E$ and $G$ are akin: the maximum value in each row of the periodic table is obtained for the group IVB TMNs. The spread between $G_R$ and $G_V$ (shaded in Fig.~\ref{fig:polycrystalline}), as well as between $E_R$ and $E_V$ is very small for the group IIIB and IVB TMNs, suggesting that the elastic properties of polycrystals of these materials will not be hugely influenced by the misorientations of individual grains. A different situation is obtained for AlN and group VB TMNs (in particular, for NbN and TaN), where the Reuss--Voigt range is quite large. As shown in Ref.~\onlinecite{Friak2008}, the ratio between the Voigt and Reuss bounds depends non-linearly on the anisotropy factor $A$. The ratio becomes particularly large when the anisotropy $A$ approaches 0, as in the case of NbN and TaN. As a consequence, these materials are expected to be strongly affected by the actual microstructures (i.e., not only by the grain orientations, but also by the shape of the grain). 

Based on an evaluation of a large experimental data set, Pugh\cite{Pugh1954} proposed that the higher (lower) the $B/G$ ratio is, the more ductile (brittle) the material is. This ratio is plotted in Fig.~\ref{fig:anisotropy}. In general, the ductility increases from IIIB to VB group (e.g., with increasing number of valence electrons and thus increasing amount of metallic bonding), and within each group from lighter to heavier elements.

\subsection{Third order elastic constants}

The methodology employing the deformation matrices $A1$--$A6$ allows also to easily estimate third order elastic constants (TOECs), by following Eq.~\ref{eq:TEOC} and relations in Table \ref{tab:combinations}. TOECs, $C_{ijk}$, appear in the Taylor series expansion of the strain energy
\begin{align}
  U&=\frac12\sum_{ij}C_{ij}\eta_i\eta_j+\frac16\sum_{ijk}C_{ijk}\eta_i\eta_j\eta_k+\dots\notag\\
   &=\frac12\sum_{ij}\left(C_{ij}+\frac13\sum_kC_{ijk}\eta_k+\dots\right)\eta_i\eta_j\ ,\label{eq:U}
\end{align}
where $\varepsilon_{\alpha\beta}=(1+\delta_{\alpha\beta})\eta_i/2$ is the relationship between components $\varepsilon_{\alpha\beta}$ of the Lagrangian strain tensor and six components $\eta_i$ of a corresponding vector in Voigt notation\cite{Lubarda1997}. According to the above equation, TOECs give corrections when  applying such large  strains that linear elasticity no longer applies. TOECs are thus useful to describe the pressure dependence of second order elastic constants, $C_{ij}$, or thermal properties of solids\cite{Cain1988}. This can be of a particular interest for thin films where residual stresses in the range of several GPa can be realized.

\begin{figure}
  \includegraphics{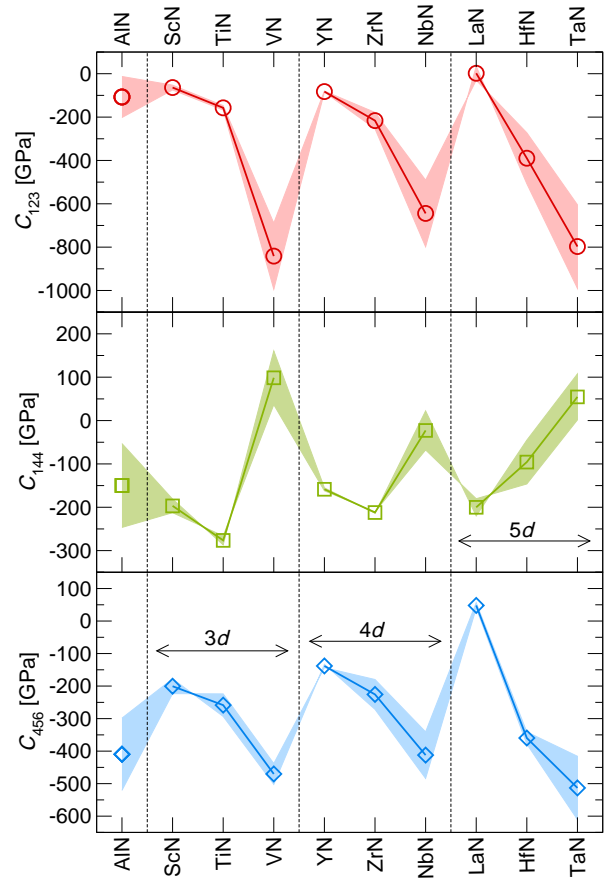}
  \caption{(Colour on-line) The polycrystalline third order elastic constants. The shaded are corresponds, for each compound, to the spread between Voight- and Reuss-type approach, while the solid line with data points represents Hill's average (Eq.~\ref{eq:Hill}). }\label{fig:polycrystalline_TOEC}
\end{figure}

The computed TOECs for the binary systems investigated in this work are summarized in Table~\ref{tab:TEOC}. For isotropic aggregates of cubic crystals, \citet{Lubarda1997} derived equations for Voigt- and Reuss-type averages of TOECs. These equations are equivalents to the elastic constants expressed by Eqs.~\ref{eq:ShearModulusVoigt}--\ref{eq:YoungsModulus}. The corresponding formulae are briefly summarised in Appendix~\ref{App.TOEC}. The three polycrystalline Voigt- and Reuss-type TOECs, $C_{123}$, $C_{144}$, and $C_{456}$ are presented in Fig.~\ref{fig:polycrystalline_TOEC} as the upper and lower boundaries of the shaded areas. These boundaries provide an estimate for the expected spread of the data and depends on the actual microstructure. The lines in Fig.~\ref{fig:polycrystalline_TOEC} represent the Hill's average
\begin{equation}
  C_{ijk,\mathrm{H}}=\frac{C_{ijk,\mathrm{V}}+C_{ijk,\mathrm{R}}}2 \label{eq:Hill}
\end{equation}
of the Voigt-type, $C_{ijk,\mathrm{V}}$, and Reuss-type, $C_{ijk,\mathrm{R}}$ TOECs. The six TOECs describing a crystal with the cubic symmetry are in the isotropic case related by
\begin{gather}
  C_{111}=C_{123}+6C_{144}+8C_{456}\ ,\\
  C_{112}=C_{123}+C_{244}\ ,\\
  C_{166}=C_{144}+2C_{456}\ .
\end{gather}

The results suggest that in each row of the periodic table, $C_{123}$ and $C_{456}$ decrease to more negative values with increasing number of valence electrons from 3 to 5. For $C_{144}$ no clear trend is observed. Since the TOECs are mostly negative, second order elastic constants get stiffer with compressive stresses while they soften under tension (compare with Eq.~\ref{eq:U}).

\begin{table}
  \begin{ruledtabular}
  \begin{tabular}{lrrrrrr}
      &  \multicolumn{1}{c}{$C_{111}$}  &  \multicolumn{1}{c}{$C_{112}$}  &  \multicolumn{1}{c}{$C_{123}$}  &  \multicolumn{1}{c}{$C_{144}$}  &  \multicolumn{1}{c}{$C_{166}$}  &  \multicolumn{1}{c}{$C_{456}$}  \\
      &  \multicolumn{1}{c}{[GPa]} &  \multicolumn{1}{c}{[GPa]} &  \multicolumn{1}{c}{[GPa]} &  \multicolumn{1}{c}{[GPa]} &  \multicolumn{1}{c}{[GPa]} &  \multicolumn{1}{c}{[GPa]}   \\\hline
    AlN  &$  -5200  $&$  -400  $&$  330  $&$  320  $&$  -850  $&$  380  $\\
    ScN  &$  -5100  $&$  -190  $&$  260  $&$  200  $&$  -330  $&$  285  $\\
    TiN  &$  -7100  $&$  -370  $&$  430  $&$  175  $&$  -475  $&$  80  $\\
    VN  &$  -8000  $&$  -400  $&$  420  $&$  541  $&$  -450  $&$  -300  $\\
    YN  &$  -4100  $&$  -160  $&$  180  $&$  180  $&$  -225  $&$  250  $\\
    ZrN  &$  -6450  $&$  -310  $&$  370  $&$  150  $&$  -370  $&$  -5  $\\
    NbN  &$  -8600  $&$  -190  $&$  115  $&$  300  $&$  -480  $&$  -235  $\\
    LaN  &$  -1200  $&$  -550  $&$  650  $&$  140  $&$  -80  $&$  180  $\\
    HfN  &$  -7050  $&$  -350  $&$  520  $&$  170  $&$  -450  $&$  -300  $\\
    TaN  &$  -9800  $&$  -20  $&$  -190  $&$  340  $&$  -600  $&$  -310  $
  \end{tabular}
  \end{ruledtabular}
  \caption{Third order elastic constants as obtained from the six deformation modes $A1$-$A6$ (Eq.~\ref{eq:TEOC}).}\label{tab:TEOC}
\end{table}

\section{Conclusions}

Calculating elastic properties using density functional theory is a powerful technique, in particular when material phases single crystals are not experimentally accessible. In this paper we provided a coherent description of the elastic behaviour of nine binary early transition metal nitrides (ScN, TiN, VN, YN, ZrN, NbN, LaN, HfN, and TaN) and AlN. These binary compounds are of high technological interest for designing materials with application-tailored properties. Single crystal elastic constants, $C_{ij}$, and directionally resolved Young's moduli, $E$, in $\langle100\rangle$, $\langle110\rangle$, and $\langle111\rangle$ directions are provided. The results clearly indicate the special position of AlN. This material has the largest Young's modulus $E$ along $\langle111\rangle$, while all group IVB and VB nitrides exhibit the largest $E$ value along $\langle100\rangle$. These trends could be rationalised by analysing the bonding characteristics of these compounds. Computing the elastic anisotropy we find that YN followed by ScN and LaN are the materials closest to the elastically isotropic behaviour. Finally, the polycrystalline elastic properties (Young's and shear modulus) were calculated. Good agreement with the rather scarce available experimental data was obtained.

\appendix

\section{Equations for the polycrystalline TOECs}\label{App.TOEC}

A detailed derivation of the equations below was given  by \citet{Lubarda1997}. Here we only summarise the final results for the reader's perusal.

The Voigt type TOECs, $C_{ijk,\mathrm{V}}$, are obtained from the single crystal elastic constants $C_{ijk}$ as
\begin{widetext}
\begin{gather}
  C_{123,\mathrm{V}}=\frac1{35}(C_{111}+18C_{112}+16C_{123}-30C_{144}-12C_{166}+16C_{456})\ ,\\
  C_{144,\mathrm{V}}=\frac1{35}(C_{111}+4C_{112}-5C_{123}+19C_{144}+2C_{166}-12C_{456})\ ,\\
  C_{456,\mathrm{V}}=\frac1{35}(C_{111}-3C_{112}+2C_{123}-9C_{144}+9C_{166}+9C_{456})\ .\label{eq:TOEC-Voigt}
\end{gather}
The Reuss-type estimates of TEOCs, $C^*_{ijk,\mathrm{R}}$, can be calculated as
\begin{gather}
  C_{456,\mathrm{R}}=\frac1{35}\left(\frac{5A}{2A+3}\right)^3\left[C_{111}-3C_{112}+2C_{123}-\frac{9}{A^2}(C_{144}-C_{166})+\frac9{A^3}C_{456}\right]\ ,\label{eq:TOEC-Reuss}\\
  C_{144,\mathrm{R}}=\frac13\left[\frac{A}{2A+3}\left[C_{111}-C_{123}+\frac3A(C_{144}+2C_{166})\right]-4C_{456,\mathrm{R}}\right]\ ,\\
  9C_{123,\mathrm{R}}+18C_{144,\mathrm{R}}+8C_{456,\mathrm{R}}=9C_{123,\mathrm{V}}+18C_{144,\mathrm{V}}+8C_{456,\mathrm{V}}\ ,
\end{gather}
\end{widetext}
where $A$ is the anisotropy ratio given by Eq.~\ref{eq:anisotropy}. It can be seen that in  case of an isotropic materials ($A=1$), the two approaches give the same results (compare e.g., Eqs.~\ref{eq:TOEC-Voigt} and \ref{eq:TOEC-Reuss}.

\section*{Acknowledgements}

The authors greatly acknowledge the financial support by the START Program (Y371) of the Austrian Science Fund (FWF).

\bibliographystyle{aipnum4-1.bst}
\bibliography{elasticity_nitrides.bib,elasticity_nitrides_extra.bib}
\end{document}